\documentclass{article}

\usepackage{natbib}

     \usepackage[preprint]{neurips_2025}

\usepackage[utf8]{inputenc} 
\usepackage[T1]{fontenc}    
\usepackage{hyperref}       
\usepackage{url}            
\usepackage{booktabs}       
\usepackage{amsfonts}       
\usepackage{nicefrac}       
\usepackage{microtype}      
\usepackage{xcolor}         
\usepackage{dirtytalk}
\usepackage{threeparttable}
\usepackage{tablefootnote}
\usepackage{adjustbox}
\usepackage{array}
\usepackage{tikz}
\usepackage{multirow}
\usepackage{listings}
\usetikzlibrary{shapes.geometric, arrows, positioning, fit, calc}

\title{Multi-IaC-Eval: Benchmarking Cloud Infrastructure as Code Across Multiple Formats}

\author{%
  Sam Davidson, Li Sun, Bhavana Bhasker, Laurent Callot, Anoop Deoras\\
  Amazon Web Services\\
  Seattle, WA \\
  \texttt{[ssdavid,bbhvna,lisunai,lcallot,adeoras]@amazon.com} \\
}

\begin{document}

\maketitle

\begin{abstract}
  Infrastructure as Code (IaC) is fundamental to modern cloud computing, enabling teams to define and manage infrastructure through machine-readable configuration files. However, different cloud service providers utilize diverse IaC formats. The lack of a standardized format requires cloud architects to be proficient in multiple IaC languages, adding complexity to cloud deployment. While Large Language Models (LLMs) show promise in automating IaC creation and maintenance, progress has been limited by the lack of comprehensive benchmarks across multiple IaC formats. We present Multi-IaC-Bench, a novel benchmark dataset for evaluating LLM-based IaC generation and mutation across AWS CloudFormation, Terraform, and Cloud Development Kit (CDK) formats. The dataset consists of triplets containing initial IaC templates, natural language modification requests, and corresponding updated templates, created through a synthetic data generation pipeline with rigorous validation. We evaluate several state-of-the-art LLMs on Multi-IaC-Bench, demonstrating that while modern LLMs can achieve high success rates (>95\%) in generating syntactically valid IaC across formats, significant challenges remain in semantic alignment and handling complex infrastructure patterns. Our ablation studies highlight the importance of prompt engineering and retry mechanisms in successful IaC generation. We release Multi-IaC-Bench to facilitate further research in AI-assisted infrastructure management and establish standardized evaluation metrics for this crucial domain.
\end{abstract}

\section{Introduction}
Infrastructure as Code (IaC), which uses machine-readable files to specify and deploy cloud computing resources, is a cornerstone of modern automation, enabling teams to define, provision, and manage their applications through code and streamlined software delivery pipelines. These pipelines support rigorous testing, enforce governance controls, and foster reliable, repeatable processes \citep{guerriero2019adoption}. By leveraging frameworks such as CloudFormation, the Cloud Development Kit (CDK), and Terraform, teams can confidently replicate applications across environments, make iterative changes, and ensure consistent infrastructure provisioning. IaC practitioners include DevOps engineers, platform engineers, cloud architects, developers, and other hybrid roles that combine software development with operational responsibilities on platforms like AWS. These professionals are engaged across the Software Development Lifecycle (SDLC), tackling tasks such as designing infrastructure solutions, authoring or modifying IaC, provisioning resources, troubleshooting deployment issues, addressing errors, re-provisioning resources, and monitoring operational infrastructure \citep{artac2017devops}.

Creating high-quality IaC for cloud deployments requires specialized domain expertise, prompting the emergence of dedicated cloud architect roles \citep{guerriero2019adoption, artac2017devops}. While software developers may focus on application logic, they often lack the contextual knowledge needed for effective cloud deployment code. The complexity of organizational requirements further necessitates that cloud architects stay up to date with evolving cloud development frameworks, often through certifications \citep{morris2016infrastructure}. Additionally, cloud providers periodically deprecate services or introduce updates that require corresponding applications to adapt, making compliance an ongoing challenge \citep{munteanu2012service}. Cloud architects must also navigate diverse integration formats offered by third-party providers, gaining proficiency in multiple infrastructure code formats. To support multi-cloud deployments, they must ensure interoperability by using common formats and deep technical knowledge \citep{guerriero2019adoption}. Unlike application code generation, infrastructure code generation demands a nuanced understanding of cloud-specific intricacies to produce scalable, reliable, and compliant IaC solutions tailored to organizational needs.

Given the complexity of creating and maintaining IaC, an obvious solution appears to be the application of Generative AI to update and generate IaC code from natural language and visual input. However, researchers have to date had only limited success in automating the creation and maintenance of IaC; specifically, the use of Large Language Models (LLMs) to generate and modify cloud infrastructure configuration files, such as AWS CloudFormation, AWS CDK, and Terraform, has been explored only to a limited degree in the academic literature. One key hurdle to the development of IaC generation pipelines is that few benchmark datasets exist to evaluate the competence of AI systems to effectively generate IaC, and those that do exist offer limited coverage of the diverse set of available IaC formats and use cases. Thus, one of the most pressing challenges is the need to create datasets of curated examples for system evaluation and model training across multiple IaC formats. We need datasets that are accurate, challenging, and representative of the distribution of user requests and data quality that we will see in a production environment.

Unfortunately, at present, available datasets are quite limited, making evaluation of LLM-generation of IaC quite challenging. Thus, creation and maintenance of IaC code is still a largely manual process. In this paper, we propose a novel benchmark dataset to evaluate the task of using natural language requests to generate and/or modify IaC templates. We harness an LLM-based synthetic data generation platform to create synthetic data by mutating templates sourced from public GitHub repositories and then describing the mutations made as a customer request. Each data point consists of three components: an initial IaC template (which may be empty), a natural language user request to create or modify one or more resources in the template, and an updated template that implements the change requested by the user.

Using the newly proposed benchmark data, we then explore the effectiveness of various LLMs at generating and mutating IaC code based on natural language input. We demonstrate that LLMs are capable of generating syntactically correct IaC files across all three formats tested, as confirmed by both appropriate linters and other static analysis tools such as Checkov. Additionally, we use an LLM judge to review the semantics of the generated outputs to determine if the LLMs we tested are able to effectively implement the changes requested in the input utterances and implement the same functionality as the reference IaC; we validate the LLM judge by conducting a human review of a sample of output IaC files. Our results show that while current models are capable of generating high-quality, user-compliant IaC in many cases, there remains significant room for improvement, especially in alignment with natural language requests. We hope that the provided dataset will facilitate additional work in IaC generation and mutation from natural language input. We release our benchmark data at \url{huggingface.co/datasets/AmazonScience/Multi-IaC-Eval}.

\section{Related work}
Although previous work in the use of LLMs to generate and mutate IaC files based on natural language requests has been limited, especially considering the utility and popularity of IaC in the DevOps community, a few researchers have explored this task. For example, \citet{xu2023cloudevalyaml} proposes a benchmark dataset for evaluation of the ability of LLMs to generate a diverse set of YAML cloud configuration files in Kubernetes format from natural language (NL) input. The dataset contains hand-selected cloud application configurations, along with hand-written NL descriptions and unit tests to test the generated configurations. The dataset covers configuration files for a variety of cloud-native applications. The benchmark dataset consists of 337 natural language description/ground-truth Kubernetes YAML/unit-test triples. The authors also provide code to efficiently benchmark generated Kubernetes files in parallel using the unit tests. Finally, they provide results on their dataset for 13 different LLMs of varying size and training techniques. They show that GPT-4-Turbo \citep{achiam2023gpt} is able to generate YAML configurations that pass all unit tests for 56.7\% of the test cases. Other members of the GPT family perform reasonably well. Open-source models do not perform as well (best performing is Llama-2-70b-chat \citep{touvron2023llama} at 8.9\% unit test pass). 

In another work, \citet{pujar2023automated} presents Ansible Wisdom, \say{a natural-language-to-Ansible YAML code generation tool, aimed at improving IT automation productivity.} The paper describes both in-context learning and fine-tuning experiments to generate NL→Ansible YAML files. No dataset is provided; however, they describe the scraping of both pre-training and fine-tuning data (pretraining data from Google BigQuery\footnote{\url{https://cloud.google.com/bigquery}}; fine-tuning data from Ansible Galaxy\footnote{\url{https://galaxy.ansible.com/}}). They use this general YAML data from BigQuery for pretraining of CodeGen \citep{nijkamp2022codegen}, followed by fine-tuning on curated Ansible data from Ansible Galaxy. Their best performing fine-tuned model correctly enforces the Ansible YAML format in 98\% of test cases, and is equivalent to the ground-truth in 70.79\% of cases. Their experiments with in-context learning on LLMs are not as successful, though they did not experiment with OpenAI's GPT or Anthropic's Claude family of models (due to the date of the work). Similarly, \citet{srivatsa2023survey} reports 56.81\% functional equivalency (that is, if the file compiles and results in the same infrastructure settings) with human-written reference files when generating Ansible-YAML files with GPT-3.5 from NL input. This again lends further credence to the ability of larger LLMs to successfully generate IaC files from NL descriptions. However, neither \citet{pujar2023automated} nor \citet{srivatsa2023survey} offer a publicly available evaluation dataset.

Recent work has expanded the benchmarking landscape for NL-to-IaC tasks, most notably with the introduction of IaC-Eval \citep{kon2024iac}. This benchmark targets LLM generation of Terraform scripts—another widely used infrastructure language—and consists of 458 human-curated scenarios spanning a wide range of AWS services and intent specifications. While quantitative results in earlier works were promising in YAML or Ansible scripts, evaluations on IaC-Eval reveal a significant performance gap: state-of-the-art LLMs such as GPT-4 achieve less than 20\% pass@1 accuracy, indicating substantial challenges in handling real-world, compositional IaC requirements. Furthermore, most benchmarks, including those above, focus exclusively on code generation from scratch, without addressing the practically important problem of mutating or incrementally updating existing IaC files in response to natural language modifications.

While this prior research is informative, it leaves substantial room for additional improvement. Of the available public datasets, only \citet{xu2023cloudevalyaml} and \citet{kon2024iac} release evaluation suites specifically for NL-to-IaC, but each focuses on a different IaC target language (Kubernetes YAML, Terraform) and neither directly addresses file mutation or round-trip NL–IaC–NL evaluation scenarios. Thus, although the recent literature demonstrates rapid progress in LLM-based IaC synthesis, the availability of comprehensive, public benchmarks for IaC remain largely open research problems.

\section{Methodology}
Given our goal of developing a benchmark dataset to evaluate the ability of diverse LLM models to mutate IaC templates from a natural language user requests, our first task was identifying which IaC formats we wished to include in the initial release of our dataset. In this initial release, we explore the IaC formats supported by Amazon Web Services (AWS) - namely AWS CloudFormation (CFN), AWS Cloud Development Kit (CDK), and HashiCorp’s widely supported Terraform (TF) format. While we limit our initial data release and evaluation to these three AWS-preferred formats, we plan to expand to additional formats and experimentation with other cloud platforms in subsequent releases of the dataset.

\subsection{Data format}
Our benchmark data consists of triplets containing an initial IaC template or repository (in the case of CDK), a natural language request to add or update one or more resources in the original IaC code, and an updated template (or repository) that implements the changes specified in the natural language request. In our dataset, we refer to these items as \say{initial}, \say{utterance}, and \say{expected}, respectively. Thus, each triplet consists of a web-sourced IaC template or repository (for CDK) along with a synthetically generated user request that is relevant to the existing IaC settings and a synthetically mutated IaC file representing the implementation of the change specified in the natural language request. The initial template and user utterance can be considered the system input to an automated natural language to IaC mutation system, while the updated IaC template implementing the user request is the expected system output. We also include instances in which the initial template contains no values to represent cases where the user either has no preexisting cloud infrastructure settings, or in which the user is requesting a from-scratch IaC generation.

\subsection{Data sources}
\label{subsec:data_sources}

We source the IaC templates used as the basis of our synthetic data generation process from three public repositories as shown in Table \ref{tab:data_sources}. The majority of our data is sourced from the \texttt{iac-model-evaluation} repository, which contains data in all three formats used in our dataset. Additional CloudFormation and Terraform data sourced from the \texttt{iac-eval} and \texttt{aws-cloudformation-templates} repositories, respectively. To improve input data quality, we filter the sourced data using static anaylsis tool; for CFN and Terraform formats, each scraped template is checked with CFNLint\footnote{\url{https://github.com/aws-cloudformation/cfn-lint}} or TFLint\footnote{\url{https://github.com/terraform-linters/tflint}}, respectively, as well as with Checkov\footnote{\url{https://www.checkov.io/}}. For all tools, we use the format-appropriate default ruleset. As our CDK data is automatically converted from CFN data (as described in Secion \ref{subsec:cdk_conversion}, below), we do not use sourced CDK data directly in the synthetic generation of evaluation triplets.

\begin{table}[h]
\centering
    \caption{IaC file data sources}
    \label{tab:data_sources}
    \begin{tabular}{|l|c|c|}
        \hline
        GitHub Repository & CFN & Terraform \\
        \hline
        iac-model-evaluation\tablefootnote{\url{https://github.com/aws-cloudformation/iac-model-evaluation}} & 39 & 223 \\
        iac-eval\tablefootnote{\url{https://github.com/autoiac-project/iac-eval}} & 0 & 10 \\
        aws-cloudformation-templates\tablefootnote{\url{https://github.com/aws-cloudformation/aws-cloudformation-templates}} & 49 & 0 \\
        \hline
    \end{tabular}
\end{table}


\subsection{Synthetic data generation}

To create quality synthetic data for the NL-to-Iac task, we first source IaC files in one of two formats - Terraform or AWS CloudFormation - as discussed in section \ref{subsec:data_sources}. We focus on these two formats due to their structure: in both cases a full cloud infrastructure stack can be defined in a a single template file. The fact that these templates can be easily represented as structured text from a single input file facilitates use of these formats as input to an LLM for mutation. Additionally, we are able to use synthetic triplets generated in CFN format to create CDK format triplets, as detailed in Section \ref{subsec:cdk_conversion}, allowing us to expand the scope of our proposed dataset. 

It is important to note that the initial IaC templates provided as input to the synthetic data pipeline are varied in complexity, and each template serves as the basis for multiple synthetic examples. In the simplest cases, the input contains an empty IaC skeleton to represent a user request to generate a completely new IaC template, while other input templates define multiple cloud resources with associated properties and security configurations, representing customers seeking to modify their complex existing cloud deployments. By using each template multiple times as input, we can generate diverse examples of how users might want to modify or extend a given infrastructure configuration. This diversity in both input complexity and generated modifications enables our pipeline to generate synthetic data that reflects a broad range of real-world infrastructure management scenarios, from initial infrastructure setup to modifications of complex existing deployments.

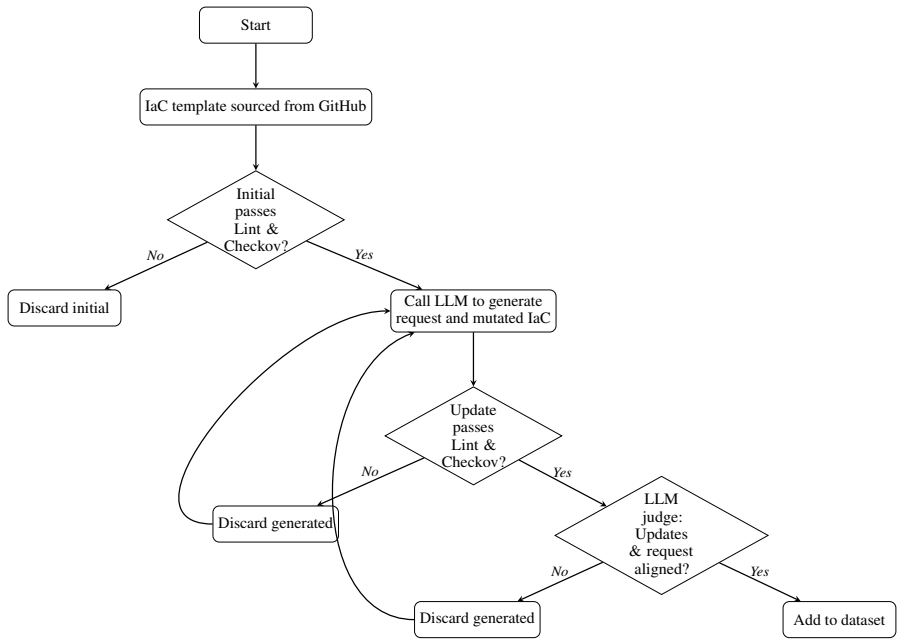
\begin{figure}[ht!]
    \centering
    \scalebox{0.6}{
    \begin{tikzpicture}[
        node distance = 1cm and 1.5cm,
        box/.style = {rectangle, rounded corners, draw, minimum width=2.5cm, minimum height=0.8cm, align=center},
        decision/.style = {diamond, draw, text width=4em, aspect=1.8, inner sep=0pt, align=center},
        arrow/.style = {->, >=stealth, thick},
        desc/.style = {font=\small\itshape, text width=3cm, align=center},
        ]
        
        \node[box] (start) {Start};
        \node[box, below=of start] (iac) {IaC template sourced from GitHub};
        \node[decision, below=of iac] (check1) {Initial passes Lint \& Checkov?};
        \node[box, below left=1cm and 2cm of check1] (discard1) {Discard initial};
        \node[box, below right=1cm and 2cm of check1] (llm) {Call LLM to generate\\ request and mutated IaC};
        \node[decision, below=1.2cm of llm] (check2) {Update passes Lint \& Checkov?};
        \node[box, below left=1cm and 2cm of check2] (discard2) {Discard generated};
        \node[decision, below right=1cm and 2cm of check2] (check3) {LLM judge: Updates \& request aligned?};
        \node[box, below right=0.8cm and 1.5cm of check3] (add) {Add to dataset};
        \node[box, below left=0.8cm and 1.5cm of check3] (discard3) {Discard generated};
        
        \draw[arrow] (start) -- (iac);
        \draw[arrow] (iac) -- (check1);
        \draw[arrow] (check1) -- node[desc, above] {No} (discard1);
        \draw[arrow] (check1) -- node[desc, above] {Yes} (llm);
        \draw[arrow] (llm) -- (check2);
        \draw[arrow] (check2) -- node[desc, above] {No} (discard2);
        \draw[arrow] (check2) -- node[desc, above] {Yes} (check3);
        \draw[arrow] (check3) -- node[desc, above] {No} (discard3);
        \draw[arrow] (check3) -- node[desc, above] {Yes} (add);
        
        \draw[arrow] (discard2) to[out=180, in=180] (llm);
        \draw[arrow] (discard3) to[out=180, in=200] (llm);
        
    \end{tikzpicture}
    }
    \caption{Flow chart of the synthetic IaC mutation triplet generation pipeline}
    \label{fig:flow-chart}
\end{figure}

For each CFN or TF file in the source dataset, we employ a two-stage process to create both the natural language "user" request and the corresponding updated IaC code, as detailed in Figure \ref{fig:flow-chart}. We first ask an LLM to review the input IaC template and determine one or more possible infrastructure change(s) that a customer might realistically request given the current IaC configuration, and to describe the change(s) in natural language. We then further prompt the LLM to generate an updated version of the IaC template that implements the change(s) requested by the fictitious user.

Additionally, because we would like to create multiple request-update pairs for a single input IaC template, we keep track of all previously generated requests for each initial IaC template in our source data. When we prompt the pipeline, we add a constraint requesting that its next output for the provided input be substantially different from the previous requests. This approach ensures that our synthetic dataset captures a diverse range of possible infrastructure modifications while avoiding redundant or overly similar examples.

We provide examples of the prompts used to generate our synthetic user request and mutated IaC templates in Appendix A. All data is generated using the Amazon Nova Pro model \citep{amazon_nova} in AWS Bedrock \citep{aws_bedrock}. We set the model temperature to 0.9 to promote output diversity, while leaving all other parameters at their default values. We additionally show the top 10 resource types covered in our generated dataset for each of the target IaC formats in Table \ref{tab:combined_dist}

\begin{table}
\centering
\small  
\caption{Top 10 most commonly used resources by IaC type in Multi-IaC-Bench}
\setlength{\tabcolsep}{2pt}  
\begin{tabular}{p{3.5cm}c|p{3.5cm}c|p{3.5cm}c}  
\toprule
CloudFormation Resource & Count & Terraform Resource & Count & CDK Resource & Count \\ 
\midrule
AWS::IAM::Role & 85 & aws\_s3\_bucket & 73 & AWS::EC2::Subnet & 95 \\
AWS::EC2::Subnet & 71 & aws\_iam\_role & 44 & AWS::S3::Bucket & 57 \\
AWS::EC2::SecurityGroup & 71 & aws\_iam\_role\_policy & 37 & AWS::SSM::Parameter & 50 \\
AWS::S3::Bucket & 57 & aws\_sfn\_state\_machine & 30 & AWS::EC2::VPC & 32 \\
AWS::EC2::SubnetRTAssoc. & 44 & aws\_sns\_topic & 30 & AWS::S3::BucketPolicy & 28 \\
AWS::CloudWatch::Alarm & 44 & kubernetes\_secret & 30 & AWS::IAM::Role & 25 \\
AWS::EC2::EIP & 34 & aws\_cloudwatch\_alarm & 28 & AWS::IAM::Policy & 23 \\
AWS::SNS::Topic & 33 & aws\_sns\_topic\_sub & 20 & AWS::EC2::SecurityGroup & 21 \\
AWS::EC2::RouteTable & 32 & aws\_vpc & 15 & AWS::DynamoDB::Table & 18 \\
AWS::EC2::Route & 32 & kubernetes\_config\_map & 15 & AWS::Budgets::Budget & 17 \\ 
\bottomrule
\end{tabular}
\label{tab:combined_dist}
\end{table}

\subsection{CDK data conversion}
\label{subsec:cdk_conversion}

The modification of IaC data in the AWS Cloud Developer Kit (CDK) format is somewhat more complex than mutating CFN or Terraform data, principly becuase CDK is provided as a code repository in which multiple files can contain various aspects of the resources defined by the IaC code. In many cases, a single stack file will contain all of the structural definitions in the CDK repository, but this is not always true, especially for more complex infrastructure stacks. As such, the use of LLMs to understand that content of IaC code in CDK format and to mutate the same requires an agentic system capable exploring the CDK repository to identify which files contain resource definitions, property values, parameters, and other relevant information, and then to mutate the appropriate files according to the natural language user request. We are currently developing a system of this type to allow natural language based mutation of CDK code; however, such a system is beyond the scope of the present paper and dataset.

To expand the scope of the present dataset to include CDK-format IaC, we take advantage of the AWS CDK Migrate utility\footnote{https://docs.aws.amazon.com/cdk/v2/guide/ref-cli-cdk-migrate.html} that converts input CloudFormation templates to level-1 CDK repositories in a deterministic, rule-based manner. To create a new dataset triple - containing an initial IaC, utterance, and expected IaC - we convert both the intital and expected CFN templates in our mutated CFN data to CDK format. We conduct this conversion on a sample of 96 of our CFN data, converting to both Python (96) and Typescript (95) formatted CDK. Thus, our final dataset contains 191 CDK triplets in two programming languages. We use the same conversion process to facilitate evaluation of CDK data in our Experiments section below.

\subsection{Data validation}

We validate the quality of our generated benchmark data in two key ways. First, we use Checkov, CFNLint, and TFLint to ensure that all source IaC templates used as the basis of IaC mutation meet standards of IaC best practices and security. We additionally ensure that all generated templates in our dataset pass these same static analysis tools with no errors. Thus, all templates in our dataset, both the original initial template and the synthetic expected template, pass Checkov and their format-specific linter. Given that our CDK data is deterministically mapped from passing initial and expected CFN pairs, we do not conduct further static analysis of the CDK repositories in our dataset. We remove all initial templates that contain lint and Checkov errors from our source dataset prior to running our pipeline. During generation, we remove any expected template that does not pass lint and Checkov along with its associated synthetic user request, and generate a new triplet from the source template.  

While static tools can ensure that our input and output IaC meets basic guidelines for quality, they cannot ensure semantic alignment between the synthetic user request and the generated target IaC template. Nor can such tools ensure that the mutated IaC output is faithful to the input IaC template, beyond the changes necessary to implement the user's request. To check these two aspects of data quality we use an LLM judge. We pass the original template, the natural language request, and the mutated template of the LLM and ask it to check if the changes made between the original and the mutated templates align with the natural language request, and if any other changes have been made beyond those needed to implement the request. Our LLM judge indicates alignment and faithfulness in 96\% of the synthetic data triplets we generated using Nova; we remove those samples that do not comply from the dataset.

As a final step to ensure that our benchmark is of high quality, a sample of 60 generated IaC mutation triplets were reviewed by human evaluators with extensive experience with various IaC formats. Like the LLM judge, our human evaluators checked that the difference between the initial and expected template aligns with the user request for each synthetic data triplet, and that the changes made do not exceed those needed to implement the request. We found that the human reviewers had a 91.6\% alignment with the LLM judge, with a Pearson's coefficient of 0.64, showing strong positive correlation. This strong alignment between the LLM judge and our human evaluators demonstrates the effectiveness of the LLM judge as a means of validating data quality.

\section{Experiments}
\label{sec:experiments}
In this section, we comprehensively evaluate various models' performance in handling Infrastructure as Code (IaC) mutations using our Multi-IaC-Bench framework. Our evaluation spans across three popular IaC formats: CloudFormation, Terraform, and CDK, providing a thorough assessment of model capabilities across different infrastructure definition paradigms.

\subsection{Performance of Various Foundation Models for CloudFormation}

We conduct extensive benchmarking experiments on 337 CloudFormation templates from our Multi-Iac-Bench dataset using three state-of-the-art foundation models: Llama 3.2 11B Instruct, Deepseek R1, and Sonnet 3.5 V2, as shown in Table \ref{tab:benchmark_combined}. Our evaluation framework employs six carefully selected metrics that can be categorized into: (1) safety and best practice metrics (CFN-Lint and Checkov pass rate), (2) distance-based metrics (edit distance), (3) efficiency metrics (average number of LLM calls per test case), and (4) semantic alignment (LLM judge score). We use an LLM judge to check the semantic alignment between generated templates and the corresponding template in our benchmark dataset. The LLM judge acts as a proxy for functional equivalence between the benchmark and generated IaC. We additionally experiment with generation incorporating a retry mechanism, in which any output that does not pass all metrics is regenerated with a new prompt that includes information about previous errors. All experiments were conducted with a temperature setting of 0.5 to balance creativity and consistency in generation.

\begin{table}[htbp]
\centering
\caption{Benchmark Results for Different IaC Formats}
\adjustbox{max width=\textwidth}{
\begin{tabular}{@{}llccccc@{}}
\toprule
Format & Base Model Name & 
\begin{tabular}[c]{@{}c@{}}Lint\\Pass Rate\end{tabular} & 
\begin{tabular}[c]{@{}c@{}}Checkov\\Pass Rate\end{tabular} & 
\begin{tabular}[c]{@{}c@{}}Number of\\LLM Calls\end{tabular} & 
\begin{tabular}[c]{@{}c@{}}LLM Judge\\Score\end{tabular} & 
\begin{tabular}[c]{@{}c@{}}Edit\\Distance\end{tabular} \\
\midrule
\multirow{3}{*}{CFN} 
& Llama 3.2 11B Instruct & 72.11\% & 89.91\% & 2.71 & 1.89 & 822 \\
& DeepSeek R1 & 91.99\% & 92.58\% & 1.79 & 2.06 & 733 \\
& Sonnet 3.5 V2 & 98.52\% & 98.81\% & 1.82 & 2.23 & 1190 \\
\midrule
\multirow{3}{*}{Terraform}
& Llama 3.2 11B Instruct & 84.80\% & 100\% & 2.73 & 2.01 & 1343 \\
& DeepSeek R1 & 98.83\% & 98.83\% & 1.81 & 2.12 & 1061 \\
& Sonnet 3.5 V2 & 100\% & 100\% & 2.1 & 2.39 & 1403 \\
\midrule
\multirow{3}{*}{CDK}
& Llama 3.2 11B Instruct & 36.65\% & 98.43\% & 3.72 & 1.31 & 3835 \\
& DeepSeek R1 & 85.34\% & 85.86\% & 1.6 & 1.65 & 3629 \\
& Sonnet 3.5 V2 & 95.81\% & 96.34\% & 1.59 & 1.75 & 3568 \\
\bottomrule
\end{tabular}}
\label{tab:benchmark_combined}
\end{table}

Our results for CloudFormation templates demonstrate varying levels of performance across the evaluated models. Sonnet 3.5 V2 achieved 98.52\% CFN-Lint and 98.81\% Checkov pass rates. DeepSeek R1 recorded 91.99\% and 92.58\% on these metrics, respectively, with an average of 1.79 LLM calls per test case. Llama 3.2 11B Instruct showed 72.11\% CFN-Lint and 89.91\% Checkov pass rates, requiring an average of 2.71 LLM calls. 

\subsection{Performance of Various Foundation Models for Terraform}

For Terraform format evaluation, we modified our prompting strategy to accommodate Terraform-specific syntax and incorporated TF-Lint and Checkov checks in the retry loop. We evaluated performance using similar metrics adapted for Terraform, including TF-Lint and Checkov pass rates, edit distance, and LLM judge score. Our benchmark comprised 171 Terraform test cases, as shown in Table.~\ref{tab:benchmark_combined}. The results for Terraform format show a similar pattern to CloudFormation in both TF-Lint and Checkov compliance. Notably, all models demonstrated strong performance in Checkov compliance, suggesting that security best practices are well-maintained across different model architectures when handling Terraform code.

\subsection{Performance of Various Foundation Models for CDK}
For CDK format evaluation, we implemented a novel two-step approach: first converting the initial CDK template to CloudFormation format, then prompting the LLM to update the converted template based on user requirements. This approach was empirically found to outperform direct CDK template modification, as it mitigates the tendency of LLMs to hallucinate complex CDK syntax. After modification, we convert the updated CloudFormation template back to CDK format. The evaluation results on 191 CDK templates are shown in Table.~\ref{tab:benchmark_combined}. Our evaluation of 191 CDK templates reveals that this conversion-based approach significantly improves generation quality. Sonnet 3.5 V2 achieved 95.81\% CFN-Lint and 96.34\% Checkov pass rates, requiring an average of 1.59 LLM calls. DeepSeek R1 recorded 85.34\% and 85.86\% on these metrics respectively, with 1.6 average calls. Llama 3.2 11B Instruct showed 36.65\% CFN-Lint and 98.43\% Checkov pass rates, with an average of 3.72 LLM calls. 

\subsection{Prompt Experiments and Ablation Studies}

To understand the impact of different prompting strategies, we conducted extensive experiments using three prompt variants: base prompt, prompt with best practice guidelines, and prompt with best practice guidelines and chain of thought. We additionally tested generation with and without our retry mechanism. Our results on CFN, Terraform, and CDK are shown in Table \ref{tbl:prompt_combined}. These experiments were conducted using Claude Sonnet 3.5 V2.

\begin{table}[htbp]
\centering
\caption{Results of Prompt Experiments for Different IaC Formats}
\adjustbox{max width=\textwidth}{
\begin{tabular}{@{}llccccc@{}}
\toprule
Format & Method & 
\begin{tabular}[c]{@{}c@{}}Lint\\Pass Rate\end{tabular} & 
\begin{tabular}[c]{@{}c@{}}Checkov\\Pass Rate\end{tabular} & 
\begin{tabular}[c]{@{}c@{}}Number of\\LLM Calls\end{tabular} & 
\begin{tabular}[c]{@{}c@{}}LLM Judge\\Score\end{tabular} & 
\begin{tabular}[c]{@{}c@{}}Edit\\Distance\end{tabular} \\
\midrule
\multirow{4}{*}{CFN} 
& Basic Prompt & 61.93\% & 56.25\% & 1.00 & 2.45 & 758 \\
& Full Prompt & 69.32\% & 51.14\% & 1.00 & 2.40 & 758 \\
& Full Prompt w/ Retry loop & 98.52\% & 98.81\% & 1.82 & 2.23 & 1777 \\
& Full Prompt w/ Retry loop and CoT & 96.02\% & 97.73\% & 1.94 & 2.24 & 1678 \\
\midrule
\multirow{4}{*}{Terraform}
& Basic Prompt & 17.54\% & 100\% & 1.00 & 2.38 & 949 \\
& Full Prompt & 17.54\% & 100\% & 1.00 & 2.43 & 1107 \\
& Full Prompt w/ Retry loop & 100\% & 100\% & 2.10 & 2.39 & 1403 \\
& Full Prompt w/ Retry loop and CoT & 100\% & 100\% & 2.06 & 2.41 & 1307 \\
\midrule
\multirow{4}{*}{CDK}
& Basic Prompt & 67.02\% & 88.48\% & 1.00 & 1.64 & 3667 \\
& Full Prompt & 75.92\% & 97.91\% & 1.00 & 1.71 & 3675 \\
& Full Prompt w/ Retry loop & 95.81\% & 96.34\% & 1.59 & 1.75 & 3568 \\
& Full Prompt w/ Retry loop and CoT & 96.86\% & 97.91\% & 1.57 & 1.80 & 3590 \\
\bottomrule
\end{tabular}}
\label{tbl:prompt_combined}
\end{table}

The results demonstrate that incorporating retry loops significantly improves compliance metrics across all IaC formats. For CloudFormation, the full prompt with retry loop achieved 98.52\% CFN-Lint and 98.81\% Checkov pass rates, a substantial improvement over the basic prompt. Similar patterns were observed in Terraform and CDK formats, though the magnitude of improvement varied. Interestingly, while Chain of Thought (CoT) prompting showed modest improvements in CDK template generation, its benefits were less pronounced for simpler formats like CloudFormation and Terraform.

The addition of best practice guidelines in prompts showed incremental improvements in initial generation quality, but the most substantial gains came from incorporating the retry loop mechanism. This suggests that iterative refinement based on specific error feedback is more effective than attempting to achieve perfect generation in a single pass.

\section{Discussion}
Our experimental results demonstrate several important findings regarding the current state of LLM-based IaC generation and mutation across multiple formats. The performance evaluation reveals varying capabilities among the tested models. Claude Sonnet 3.5 v2 achieved CFN-Lint and Checkov pass rates of 98.52\% and 98.81\% for CloudFormation, respectively. Llama 3.2 recorded pass rates of 72.11\% and 89.91\% on these metrics, while DeepSeek R1 showed results of 91.99\% and 92.58\%. This pattern was also observed in the Terraform format, where models demonstrated different levels of compliance with validation tools. The performance extended to the more challenging CDK format, with each model showing distinct capabilities in handling this complex format.

The impact of prompt engineering emerges as a crucial factor in successful IaC generation. Our ablation studies show that incorporating best practice guidelines and implementing retry loops significantly improves performance across all formats. For CloudFormation, enhancing the basic prompt with these additions improved CFN-Lint pass rates from 61.93\% to 98.52\%. Similar dramatic improvements were observed in Terraform (17.54\% to 100\%) and CDK (67.02\% to 95.81\%). Interestingly, while the addition of chain-of-thought reasoning showed minimal impact on performance metrics, it did slightly increase the number of required LLM calls, suggesting that simpler, more direct prompting strategies may be more efficient for IaC generation tasks.

Analysis of resource distribution in our dataset reveals comprehensive coverage of commonly used AWS services across all three formats, with some notable patterns emerging. In CloudFormation, IAM roles, EC2 resources, and S3 buckets dominate, reflecting typical enterprise cloud infrastructure requirements. Terraform shows a similar distribution but with higher representation of state machine and Kubernetes resources, indicating its popular use in container orchestration scenarios. The CDK resource distribution closely mirrors CloudFormation, as expected given our conversion methodology, though this similarity also highlights one of our study's limitations.

Several important limitations of our current work should be acknowledged. First, while our dataset provides broad coverage of AWS services, it may not fully capture the complexity of enterprise-scale cloud infrastructure deployments. Second, our evaluation metrics, though comprehensive in terms of syntactic correctness and basic semantic alignment, do not include actual deployment testing, which would provide additional validation of the generated IaC's practical utility. Third, our CDK evaluation methodology, relying on conversion from CloudFormation, may not fully represent the unique features and patterns of native CDK development. These limitations suggest valuable directions for future research, including the development of more sophisticated evaluation metrics and the expansion of native CDK examples in the dataset.

The number of LLM calls required for successful generation remained relatively consistent across models for both CloudFormation (1.79-2.71 calls) and Terraform (1.81-2.73 calls), indicating that our retry mechanisms effectively handle initial generation failures. The LLM judge scores also showed consistent patterns, with Sonnet achieving the highest semantic alignment scores across all formats (2.23-2.39). These results suggest that while current LLM technology can effectively generate and modify IaC, there remains room for improvement in reducing the need for multiple generation attempts and increasing semantic accuracy on the first try.

Our findings have significant implications for the future of DevOps and cloud infrastructure management. The high success rates achieved by current LLMs, particularly in CloudFormation and Terraform formats, suggest that automated IaC generation and modification is becoming increasingly viable for production use. However, the performance variations across formats and the continuing need for retry mechanisms indicate that careful system design and robust validation processes remain essential. As LLM capabilities continue to improve, we anticipate that these tools will become increasingly valuable for automating infrastructure management tasks, though human oversight and validation will likely remain important for the foreseeable future.

\section{Conclusion}

This paper introduces Multi-IaC-Bench, a novel benchmark dataset for evaluating LLM-based Infrastructure as Code generation and mutation across multiple formats. Our comprehensive dataset covers CloudFormation, Terraform, and CDK formats, with careful attention to both syntactic correctness and semantic alignment. We provide a robust evaluation framework incorporating multiple validation methods, including static analysis tools and LLM-based semantic evaluation, enabling thorough assessment of IaC generation capabilities.

Our experimental results demonstrate that current LLMs can achieve high success rates in IaC generation across formats, with CFN-Lint and Checkov pass rates exceeding 95\% for all three formats tested. However, significant challenges remain, particularly in handling more complex IaC structures. The performance gap between formats suggests that additional work is needed to improve the handling of sophisticated infrastructure patterns. Our ablation studies also highlight the crucial role of prompt engineering and retry mechanisms in achieving reliable IaC generation.

Several promising directions for future work emerge from our findings. First, expanding the dataset to cover additional IaC formats (Pulumi, Ansible) would increase its utility for multi-cloud scenarios. Second, developing more sophisticated semantic evaluation methods and incorporating deployment testing would provide more comprehensive validation of generated IaC. Third, creating specialized models for IaC generation with improved handling of complex infrastructure patterns could address current performance limitations. 

We believe Multi-IaC-Bench provides a valuable foundation for future research in AI-assisted cloud infrastructure management, and we hope it will facilitate continued progress in this important field. By establishing a standardized benchmark for evaluating IaC generation capabilities, we aim to accelerate the development of more effective and reliable automated infrastructure management solutions.

\bibliography{ref}

\begin{thebibliography}{13}
\providecommand{\natexlab}[1]{#1}
\providecommand{\url}[1]{\texttt{#1}}
\expandafter\ifx\csname urlstyle\endcsname\relax
  \providecommand{\doi}[1]{doi: #1}\else
  \providecommand{\doi}{doi: \begingroup \urlstyle{rm}\Url}\fi

\bibitem[Achiam et~al.(2023)Achiam, Adler, Agarwal, Ahmad, Akkaya, Aleman, Almeida, Altenschmidt, Altman, Anadkat, et~al.]{achiam2023gpt}
Josh Achiam, Steven Adler, Sandhini Agarwal, Lama Ahmad, Ilge Akkaya, Florencia~Leoni Aleman, Diogo Almeida, Janko Altenschmidt, Sam Altman, Shyamal Anadkat, et~al.
\newblock Gpt-4 technical report.
\newblock \emph{arXiv preprint arXiv:2303.08774}, 2023.

\bibitem[Artac et~al.(2017)Artac, Borovssak, Di~Nitto, Guerriero, and Tamburri]{artac2017devops}
Matej Artac, Tadej Borovssak, Elisabetta Di~Nitto, Michele Guerriero, and Damian~Andrew Tamburri.
\newblock Devops: introducing infrastructure-as-code.
\newblock In \emph{2017 IEEE/ACM 39th International Conference on Software Engineering Companion (ICSE-C)}, pages 497--498. IEEE, 2017.

\bibitem[{AWS}(2023)]{aws_bedrock}
{AWS}.
\newblock Amazon bedrock - foundation models and generative ai - aws, 2023.
\newblock URL \url{https://aws.amazon.com/bedrock/}.
\newblock Accessed: April 28, 2025.

\bibitem[{AWS}(2025)]{amazon_nova}
{AWS}.
\newblock Amazon nova foundation models, 2025.
\newblock URL \url{https://aws.amazon.com/ai/generative-ai/nova/}.
\newblock Accessed: April 28, 2025.

\bibitem[Guerriero et~al.(2019)Guerriero, Garriga, Tamburri, and Palomba]{guerriero2019adoption}
Michele Guerriero, Martin Garriga, Damian~A Tamburri, and Fabio Palomba.
\newblock Adoption, support, and challenges of infrastructure-as-code: Insights from industry.
\newblock In \emph{2019 IEEE International conference on software maintenance and evolution (ICSME)}, pages 580--589. IEEE, 2019.

\bibitem[Kon et~al.(2024)Kon, Liu, Qiu, Fan, He, Lin, Zhang, Park, Elengikal, Kang, et~al.]{kon2024iac}
Patrick~T Kon, Jiachen Liu, Yiming Qiu, Weijun Fan, Ting He, Lei Lin, Haoran Zhang, Owen~M Park, George~S Elengikal, Yuxin Kang, et~al.
\newblock Iac-eval: A code generation benchmark for cloud infrastructure-as-code programs.
\newblock \emph{Advances in Neural Information Processing Systems}, 37:\penalty0 134488--134506, 2024.

\bibitem[Morris(2016)]{morris2016infrastructure}
Kief Morris.
\newblock \emph{Infrastructure as code: managing servers in the cloud}.
\newblock " O'Reilly Media, Inc.", 2016.

\bibitem[Munteanu et~al.(2012)Munteanu, Fortis, and Negru]{munteanu2012service}
Victor~Ion Munteanu, Teodor-Florin Fortis, and Viorel Negru.
\newblock Service lifecycle in the cloud environment.
\newblock In \emph{2012 14th International Symposium on Symbolic and Numeric Algorithms for Scientific Computing}, pages 457--464. IEEE, 2012.

\bibitem[Nijkamp et~al.(2023)Nijkamp, Pang, Hayashi, Tu, Wang, Zhou, Savarese, and Xiong]{nijkamp2022codegen}
Erik Nijkamp, Bo~Pang, Hiroaki Hayashi, Lifu Tu, Huan Wang, Yingbo Zhou, Silvio Savarese, and Caiming Xiong.
\newblock Codegen: An open large language model for code with multi-turn program synthesis.
\newblock \emph{ICLR}, 2023.

\bibitem[Pujar et~al.(2023)Pujar, Buratti, Guo, Dupuis, Lewis, Suneja, Sood, Nalawade, Jones, Morari, et~al.]{pujar2023automated}
Saurabh Pujar, Luca Buratti, Xiaojie Guo, Nicolas Dupuis, Burn Lewis, Sahil Suneja, Atin Sood, Ganesh Nalawade, Matt Jones, Alessandro Morari, et~al.
\newblock Automated code generation for information technology tasks in yaml through large language models.
\newblock In \emph{2023 60th ACM/IEEE Design Automation Conference (DAC)}, pages 1--4. IEEE, 2023.

\bibitem[Srivatsa et~al.(2023)Srivatsa, Mukhopadhyay, Katrapati, and Shrivastava]{srivatsa2023survey}
Kalahasti~Ganesh Srivatsa, Sabyasachi Mukhopadhyay, Ganesh Katrapati, and Manish Shrivastava.
\newblock A survey of using large language models for generating infrastructure as code.
\newblock In \emph{Proceedings of the 20th International Conference on Natural Language Processing (ICON)}, pages 523--533, 2023.

\bibitem[Touvron et~al.(2023)Touvron, Martin, Stone, Albert, Almahairi, Babaei, Bashlykov, Batra, Bhargava, Bhosale, et~al.]{touvron2023llama}
Hugo Touvron, Louis Martin, Kevin Stone, Peter Albert, Amjad Almahairi, Yasmine Babaei, Nikolay Bashlykov, Soumya Batra, Prajjwal Bhargava, Shruti Bhosale, et~al.
\newblock Llama 2: Open foundation and fine-tuned chat models.
\newblock \emph{arXiv preprint arXiv:2307.09288}, 2023.

\bibitem[Xu et~al.()Xu, Chen, Zhang, Lin, Hu, Ma, Lu, Du, Mao, Zhai, et~al.]{xu2023cloudevalyaml}
Yifei Xu, Yuning Chen, Xumiao Zhang, Xianshang Lin, Pan Hu, Yunfei Ma, Songwu Lu, Wan Du, Z~Morley Mao, Ennan Zhai, et~al.
\newblock Cloudevalyaml: A realistic and scalable benchmark for cloud configuration generation.

\end{thebibliography}
\bibliographystyle{plainnat}

\newpage
\appendix
\label{appendix}
\section{Prompts Used In This Study}

\lstset{
    breaklines=true,
    breakatwhitespace=true,
    columns=flexible,
    basicstyle=\small\ttfamily
}

\subsection{Example prompt used in synthetic data generation pipeline}

\begin{lstlisting}
You are an AI assistant tasked with reviewing customer Cloudformation templates, creating requests to modify the current settings,
and implementing the requested changes. Here's what you need to do:

1. First, you will be provided with a Cloudformation file describing cloud infrastructure settings:

Here is the Cloudformation configuration you will be working with:

<cloudformation>
{$CLOUDFORMATION}
</cloudformation>

Here are the previous changes made:

<previous_changes>
{$PREVIOUS}
</previous_changes>

2. Your task is to:
a) Determine one substantial change or addition to the Cloudformation configuration that a customer might request. Choose changes that are varied (choose both common and less common changes)
b) Make sure the change is substantially different from the previous changes listed above.
c) Create a natural language customer request that requests the change you envisioned.
d) Update the Cloudformation file to implement the change.

3. Analyze the Cloudformation structure carefully. Understand the different components and their
relationships.

4. Choose one substantial change or addition to make to the Cloudformation. This could involve:
- Adding a new resource
- Modifying an existing resource significantly
- Changing a critical setting that affects the overall infrastructure
- Making sure that the change you select is not the most obvious change - vary your changes to less common resource types.

Make only one change, but ensure it's meaningful and would have a substantial impact on the infrastructure. Also, make sure you vary the types of changes you make. Don't always make the most obvious change. Make changes over diverse less common resources.

5. After deciding on the change, imagine a natural language customer request that would prompt this modification. Then, write an summary of the all changes requested by the customer, in first person, such as "Add x ..." or "Replace y ...".

6. Implement the requested change in the Cloudformation template.

7. Provide your response in the following format:

<modified_cloudformation>
[Insert the modified Cloudformation here, with your one substantial change.]
</modified_cloudformation>

<customer_request>
[Insert the natural language customer request here]
</customer_request>

<explanation>
[Briefly explain the change you made and why you chose it]
</explanation>

Remember, make only one substantial change to the Cloudformation, create a realistic customer request for that change, and provide an explanation for your choice. Make sure you output the full body of the modified Cloudformation file - don't truncate the output to save space.
\end{lstlisting}

\subsection{Prompts Used for CloudFormation experiments}

Basic Prompt: 
\begin{lstlisting}
Here is the initial CloudFormation template:
<initial_template>
{initial_template}
</initial_template>

The user has made the following request:
<request>
{request}
</request>

When you have completed the modifications, provide your updated template in the following format:
<updated_template>
(Your entire updated CloudFormation template goes here)
</updated_template>
\end{lstlisting}

Full Prompt:
\begin{lstlisting}
You are an expert infrastructure-as-code developer tasked with generating or modifying a CloudFormation template to meet a user's request. You will be provided with an initial template and a specific request. Your job is to update the template to fulfill the request in the simplest way possible while maintaining best practices and following AWS CloudFormation standards. Cloud formation template may or may not include the resources requested. If the template does not include the resources requested, you can add the resource to the original template. Again, try to make your output as simple as possible - avoid adding resources not requested by the user or strictly required to make the template deployable or pass checkov and CFNLint. Keep it simple!

Here is the initial CloudFormation template:
<initial_template>
{initial_template}
</initial_template>
The user has made the following request:
<request>
{request}
</request>

Approach the task as follows:
1. Analyze the initial template and the user's request.
2. Identify the necessary changes or additions to fulfill the request. You should only try to fulfill the request, and nothing else.
3. Make the required modifications while adhering to the guidelines.
4. Ensure all resources are properly configured and linked.
5. Maintain a logical structure and use clear, descriptive names for resources.
Return the updated template including the parts of the original template that remains unchanged. 
When you have completed the modifications, provide your updated template in the following format:
<updated_template>
(Your entire updated CloudFormation template goes here)
</updated_template>
Remember to maintain the YAML format of the CloudFormation template and ensure that all indentation is correct. Your goal is to produce a simple, functional, and syntactically correct CloudFormation template that fulfills the user's request. Always enclose the template within the <updated_template> </updated_template> tags.
\end{lstlisting}

\subsubsection{Prompt used in retry loop}
\begin{lstlisting}
The generated template is enclosed within <updated_template> </updated_template> tags. 
    The error messages from CFN lint is enclosed within <cfn_lint> </cfn_lint> tags. The checkov messages are enclosed within <checkov> </checkov> tags. 
    Your task is to review the updated template and error messages to generate the template that resolves the error. 
    <updated_template> 
    {target_template}
    </updated_template> 
    <cfn_lint> 
    {lint}
    </cfn_lint>
    <checkov> 
    {checkov}
    </checkov>

    The updated template should adhere to the following guidelines:
    1. For S3 bucket, do not reference itself in DestinationBucketName of LoggingConfiguration. For example,
<incorrect_template>
AccessLogsBucket:
    Type: AWS::S3::Bucket
    Properties:
      VersioningConfiguration:
        Status: Enabled
      PublicAccessBlockConfiguration:
        BlockPublicAcls: true
        BlockPublicPolicy: true
        IgnorePublicAcls: true
        RestrictPublicBuckets: true
      OwnershipControls:
        Rules:
          - ObjectOwnership: BucketOwnerPreferred
      LoggingConfiguration:
        DestinationBucketName: !Ref AccessLogsBucket
        LogFilePrefix: access-logs-bucket-logs/
    UpdateReplacePolicy: Retain
    DeletionPolicy: Retain
</incorrect_template>
is not correct, because the DestinationBucketName refers to AccessLogsBucket itself. Instead, you should write
<corrected_template>
AccessLogsBucket:
    Type: AWS::S3::Bucket
    Properties:
      VersioningConfiguration:
        Status: Enabled
      PublicAccessBlockConfiguration:
        BlockPublicAcls: true
        BlockPublicPolicy: true
        IgnorePublicAcls: true
        RestrictPublicBuckets: true
      OwnershipControls:
        Rules:
          - ObjectOwnership: BucketOwnerPreferred
      LoggingConfiguration:
        LogFilePrefix: access-logs-bucket-logs/
    UpdateReplacePolicy: Retain
    DeletionPolicy: Retain
</corrected_template>
    2. For access logging bucket, enable access logging and store the log in itself, do not store logs in another bucket. For example,
<incorrect_template>
  LoggingBucket:
    Type: AWS::S3::Bucket
    Properties:
      OwnershipControls:
        Rules:
          - ObjectOwnership: BucketOwnerPreferred
      PublicAccessBlockConfiguration:
        BlockPublicAcls: true
        BlockPublicPolicy: true
        IgnorePublicAcls: true
        RestrictPublicBuckets: true
      BucketEncryption:
        ServerSideEncryptionConfiguration:
          - ServerSideEncryptionByDefault:
              SSEAlgorithm: AES256
      VersioningConfiguration:
        Status: Enabled
</incorrect_template>
is not correct, because the access logging is not enabled. And 
<incorrect_template>
  LoggingBucket:
    Type: AWS::S3::Bucket
    Properties:
      OwnershipControls:
        Rules:
          - ObjectOwnership: BucketOwnerPreferred
      PublicAccessBlockConfiguration:
        BlockPublicAcls: true
        BlockPublicPolicy: true
        IgnorePublicAcls: true
        RestrictPublicBuckets: true
      BucketEncryption:
        ServerSideEncryptionConfiguration:
          - ServerSideEncryptionByDefault:
              SSEAlgorithm: AES256
      VersioningConfiguration:
        Status: Enabled
      LoggingConfiguration:
        DestinationBucketName: !Ref LoggingBucketforLog
        LogFilePrefix: access-logs-bucket-logs/
</incorrect_template>
is not correct, because it stores the log of logging bucket in another bucket. Instead, you should write
<corrected_template>
  LoggingBucket:
    Type: AWS::S3::Bucket
    Properties:
      OwnershipControls:
        Rules:
          - ObjectOwnership: BucketOwnerPreferred
      PublicAccessBlockConfiguration:
        BlockPublicAcls: true
        BlockPublicPolicy: true
        IgnorePublicAcls: true
        RestrictPublicBuckets: true
      BucketEncryption:
        ServerSideEncryptionConfiguration:
          - ServerSideEncryptionByDefault:
              SSEAlgorithm: AES256
      VersioningConfiguration:
        Status: Enabled
      LoggingConfiguration:
        LogFilePrefix: 'logging-bucket-logs/'
</corrected_template>
    Remember to maintain the YAML format of the CloudFormation template and ensure that all indentation is correct. Your goal is to produce a fully functional and syntactically correct CloudFormation template that meets the user's requirements. Always enclose the template within the <updated_template> </updated_template> tags.  
\end{lstlisting}

\subsubsection{Chain-of-thought Prompt}
\begin{lstlisting}
You are an expert infrastructure-as-code developer tasked with generating or modifying a CloudFormation template to meet a user's request. You will be provided with an initial template and a specific request. Your job is to update the template to fulfill the request in the simplest way possible while maintaining best practices and following AWS CloudFormation standards. Cloud formation template may or may not include the resources requested. If the template does not include the resources requested, you can add the resource to the original template. Again, try to make your output as simple as possible - avoid adding resources not requested by the user or strictly required to make the template deployable or pass checkov and CFNLint. Keep it simple!

Let's review the IAC expert mutating the template 

Consider an utterance: Increase my budget Budget1 to 1000 
Let's analyze this utterance: 
Intent: Increase budget to 1000
Physical id : Budget1 
Let's review the initial template : 
<initial_template> 
Resources:
  Budget:
    Type: AWS::Budgets::Budget
    Properties:
      Budget:
        BudgetLimit:
          Amount: 100
          Unit: USD
        TimeUnit: MONTHLY
        BudgetType: COST
</initial_template> 
Analyze the intial template: 
Logical ID to change: Budget 
Current Amount : 100 
Amount to be updated: 1000 
Generating the target: 
<updated_template> 
Resources:
  Budget:
    Type: AWS::Budgets::Budget
    Properties:
      Budget:
        BudgetLimit:
          Amount: 1000
          Unit: USD
        TimeUnit: MONTHLY
        BudgetType: COST
</updated_template> 

Now you are given an initial template and the user request.
Here is the initial CloudFormation template:
<initial_template>
{initial_template}
</initial_template>
The user has made the following request:
<request>
{request}
</request>

Approach the task as follows:
1. Analyze the initial template and the user's request.
2. Identify the necessary changes or additions to fulfill the request. You should only try to fulfill the request, and nothing else.
3. Make the required modifications while adhering to the guidelines.
4. Ensure all resources are properly configured and linked.
5. Maintain a logical structure and use clear, descriptive names for resources.
Return the updated template including the parts of the original template that remains unchanged. 
When you have completed the modifications, provide your updated template in the following format:
<updated_template>
(Your entire updated CloudFormation template goes here)
</updated_template>
Remember to maintain the YAML format of the CloudFormation template and ensure that all indentation is correct. Your goal is to produce a simple, functional, and syntactically correct CloudFormation template that fulfills the user's request. Always enclose the template within the <updated_template> </updated_template> tags.
\end{lstlisting}

\subsection{Prompts Used for Terraform experiments}
Basic Prompt: 
\begin{lstlisting}
Here is the initial Terraform template:
<initial_template>
{initial_template}
</initial_template>
The user has made the following request:
<request>
{request}
</request>
When you have completed the modifications, provide your updated template in the following format:
<updated_template>
(Your entire updated Terraform template goes here)
</updated_template>
\end{lstlisting}

Full Prompt:
\begin{lstlisting}
You are an expert infrastructure-as-code developer tasked with generating or modifying a Terraform template to meet a user's request. You will be provided with an initial template and a specific request. Your job is to update the template to fulfill the request while maintaining best practices and following Terraform standards. Terraform template may or may not include the resources requested. If the template does not include the resources requested, you can add the resource to the original template. 
Here is the initial Terraform template:
<initial_template>
{initial_template}
</initial_template>
The user has made the following request:
<request>
{request}
</request>

Approach the task as follows:
1. Analyze the initial template and the user's request.
2. Identify the necessary changes or additions to fulfill the request.
3. Make the required modifications while adhering to the guidelines.
4. Ensure all resources are properly configured and linked.
5. Maintain a logical structure and use clear, descriptive names for resources.
Return the updated template including the parts of the original template that remains unchanged. 
When you have completed the modifications, provide your updated template in the following format:
<updated_template>
(Your entire updated Terraform template goes here)
</updated_template>

Your goal is to produce a fully functional and syntactically correct Terraform
template that meets the user's requirements. Always enclose the template within the <updated_template> </updated_template> tags.
\end{lstlisting}

\subsubsection{Prompt used in the retry loop}
\begin{lstlisting}
The generated template is enclosed within <updated_template> </updated_template> tags. 
    The error messages from TF-Lint is enclosed within <tf_lint> </tf_lint> tags. 
    The checkov messages are enclosed within <checkov></checkov> tags. 
    Your task is to review the updated template and error messages to generate the template that 
    resolves the error. 
    <updated_template> 
    {target_template}
    </updated_template>
    <tf_lint> 
    {lint}
    </tf_lint> 
    <checkov> 
    {checkov}
    </checkov>

    Remember to maintain the format of the Terraform template and ensure that all indentation is correct. Your goal is to produce a fully functional and syntactically correct Terraform template that meets the user's requirements. Always enclose the template within the <updated_template> </updated_template> tags. 
\end{lstlisting}

\subsubsection{Chain-of-Thought Prompt}
\begin{lstlisting}
You are an expert infrastructure-as-code developer tasked with generating or modifying a Terraform template to meet a user's request. You will be provided with an initial template and a specific request. Your job is to update the template to fulfill the request while maintaining best practices and following Terraform standards. Terraform template may or may not include  the resources requested. If the template does not include the resources requested, you can add the resource to the original template. 

Let's review the IAC expert mutating the template 

Consider an utterance: Increase my budget Budget1 to 1000 
Let's analyze this utterance: 
Intent: Increase budget to 1000
Physical id : Budget1 
Let's review the initial template : 
<initial_template> 
resource "aws_budgets_budget" "main" {{
  name              = "monthly-budget"
  budget_type       = "COST"
  limit_amount      = "100"
  limit_unit        = "USD"
  time_unit         = "MONTHLY"
}}
</initial_template> 
Analyze the intial template: 
Logical ID to change: Budget 
Current Amount : 100 
Amount to be updated: 1000 
Generating the target: 
<updated_template> 
resource "aws_budgets_budget" "main" {{
  name              = "monthly-budget"
  budget_type       = "COST"
  limit_amount      = "1000"
  limit_unit        = "USD"
  time_unit         = "MONTHLY"
}}
</updated_template> 

Now you are given an initial template and the user request.
Here is the initial Terraform template:
<initial_template>
{initial_template}
</initial_template>
The user has made the following request:
<request>
{request}
</request>

Approach the task as follows:
1. Analyze the initial template and the user's request.
2. Identify the necessary changes or additions to fulfill the request.
3. Make the required modifications while adhering to the guidelines.
4. Ensure all resources are properly configured and linked.
5. Maintain a logical structure and use clear, descriptive names for resources.
Return the updated template including the parts of the original template that remains unchanged. When you have completed the modifications, provide your updated template in the following format:
<updated_template>
(Your entire updated Terraform template goes here)
</updated_template>

Your goal is to produce a fully functional and syntactically correct Terraform
template that meets the user's requirements. Always enclose the template within the <updated_template> </updated_template> tags.
\end{lstlisting}

\subsection{Prompts Used for CDK experiments}
\subsubsection{Basic Prompt}
\begin{lstlisting}
Here is the initial CDK template:
<initial_template>
{initial_template}
</initial_template}
The user has made the following request:
<request>
{request}
</request>
When you have completed the modifications, provide your updated template in the following format:
<updated_template>
(Your entire updated CDK template goes here)
</updated_template>
\end{lstlisting}

\subsubsection{Full Prompt}
\begin{lstlisting}
You are an expert infrastructure-as-code developer tasked with generating or modifying a CloudFormation template to meet a user's request. You will be provided with an initial CDK stack and a specific request. Your job is to update the template to fulfill the request while maintaining best practices and following AWS CDK standards. The initial stack may or may not include  the resources requested. If the stack does not include the resources requested, you can add the resource to the original stack.
Here is the initial CDK stack:
<initial_CDK>
{initial_template}
</initial_CDK>
The user has made the following request:
<request>
{request}
</request>

Approach the task as follows:
1. Analyze the initial stack and the user's request.
2. Identify the necessary changes or additions to fulfill the request.
3. Make the required modifications while adhering to the guidelines.
4. Ensure all resources are properly configured and linked.
5. Maintain a logical structure and use clear, descriptive names for resources.
Return the updated stack including the parts of the original stack that remains unchanged. When you have completed the modifications, provide your updated stack in the following format:
<updated_CDK>
(Your entire updated CDK stack goes here)
</updated_CDK>

Your goal is to produce a fully functional and syntactically correct CDK stack that meets the user's requirements. Always enclose the stack within the <updated_CDK> </updated_CDK> tags.
\end{lstlisting}

\subsubsection{Prompt used in retry loop}
\begin{lstlisting}
The generated template is enclosed within <updated_template> </updated_template> tags. 
    The error messages from CFN Lint on is enclosed within <cfn_lint> </cfn_lint> tags. The 
    checkov messages are enclosed within <checkov> </checkov> tags. 
    Your task is to review the updated template and error messages to generate the CDK template that 
    resolves the error. 
    <updated_CDK> 
    {target_template}
    </updated_CDK> 
    <cfn_lint> 
    {lint}
    </cfn_lint>
    <checkov> 
    {checkov}
    </checkov>

    Remember to maintain the format of the CDK stack and ensure that all indentation is correct. Your goal is to produce a fully functional and syntactically correct CDK stack that meets the user's requirements. Always enclose the template within the <updated_CDK> </updated_CDK> tags.  
\end{lstlisting}

\subsubsection{Chain-of-Thought Prompt}
\begin{lstlisting}
You are an expert infrastructure-as-code developer tasked with generating or modifying a CloudFormation template to meet a user's request. You will be provided with an initial CDK stack and a specific request. Your job is to update the template to fulfill the request while maintaining best practices and following AWS CDK standards. The initial stack may or may not include  the resources requested. If the stack does not include the resources requested, you can add the resource to the original stack.

Let's review the IAC expert mutating the template 

Consider an utterance: Increase my budget Budget1 to 1000 
Let's analyze this utterance: 
Intent: Increase budget to 1000
Physical id : Budget1 
Let's review the initial template : 
<initial_template> 
from aws_cdk import (
    Stack,
    aws_budgets as budgets
)
from constructs import Construct

class InitialStack(Stack):
    def __init__(self, scope: Construct, construct_id: str, **kwargs) -> None:
        super().__init__(scope, construct_id, **kwargs)

        # Create a $100 monthly cost budget
        budget = budgets.CfnBudget(
            self, 
            "Budget",
            budget=budgets.CfnBudget.BudgetDataProperty(
                budget_limit=budgets.CfnBudget.SpendProperty(
                    amount=100,
                    unit="USD"
                ),
                time_unit="MONTHLY",
                budget_type="COST"
            )
        )
</initial_template> 
Analyze the intial template: 
Logical ID to change: Budget 
Current Amount : 100 
Amount to be updated: 1000 
Generating the target: 
<updated_template> 
from aws_cdk import (
    Stack,
    aws_budgets as budgets
)
from constructs import Construct

class InitialStack(Stack):
    def __init__(self, scope: Construct, construct_id: str, **kwargs) -> None:
        super().__init__(scope, construct_id, **kwargs)

        # Create a $1000 monthly cost budget
        budget = budgets.CfnBudget(
            self, 
            "Budget",
            budget=budgets.CfnBudget.BudgetDataProperty(
                budget_limit=budgets.CfnBudget.SpendProperty(
                    amount=1000,
                    unit="USD"
                ),
                time_unit="MONTHLY",
                budget_type="COST"
            )
        )
</updated_template> 

Now you are given an initial template and the user request.
Here is the initial CDK stack:
<initial_CDK>
{initial_template}
</initial_CDK>
The user has made the following request:
<request>
{request}
</request>

Approach the task as follows:
1. Analyze the initial stack and the user's request.
2. Identify the necessary changes or additions to fulfill the request.
3. Make the required modifications while adhering to the guidelines.
4. Ensure all resources are properly configured and linked.
5. Maintain a logical structure and use clear, descriptive names for resources.
Return the updated stack including the parts of the original stack that remains unchanged. When you have completed the modifications, provide your updated stack in the following format:
<updated_CDK>
(Your entire updated CDK stack goes here)
</updated_CDK>

Your goal is to produce a fully functional and syntactically correct CDK stack that meets the user's requirements. Always enclose the stack within the <updated_CDK> </updated_CDK> tags.
\end{lstlisting}


\newpage
\section*{NeurIPS Paper Checklist}

\begin{enumerate}

\item {\bf Claims}
    \item[] Question: Do the main claims made in the abstract and introduction accurately reflect the paper's contributions and scope?
    \item[] Answer: \answerYes{} 
    \item[] Justification: Our abstract and intro accurately describe the paper's goals, methodology, and results.
    \item[] Guidelines:
    \begin{itemize}
        \item The answer NA means that the abstract and introduction do not include the claims made in the paper.
        \item The abstract and/or introduction should clearly state the claims made, including the contributions made in the paper and important assumptions and limitations. A No or NA answer to this question will not be perceived well by the reviewers. 
        \item The claims made should match theoretical and experimental results, and reflect how much the results can be expected to generalize to other settings. 
        \item It is fine to include aspirational goals as motivation as long as it is clear that these goals are not attained by the paper. 
    \end{itemize}

\item {\bf Limitations}
    \item[] Question: Does the paper discuss the limitations of the work performed by the authors?
    \item[] Answer: \answerYes{} 
    \item[] Justification: We discuss limitations in our Discussion section
    \item[] Guidelines:
    \begin{itemize}
        \item The answer NA means that the paper has no limitation while the answer No means that the paper has limitations, but those are not discussed in the paper. 
        \item The authors are encouraged to create a separate "Limitations" section in their paper.
        \item The paper should point out any strong assumptions and how robust the results are to violations of these assumptions (e.g., independence assumptions, noiseless settings, model well-specification, asymptotic approximations only holding locally). The authors should reflect on how these assumptions might be violated in practice and what the implications would be.
        \item The authors should reflect on the scope of the claims made, e.g., if the approach was only tested on a few datasets or with a few runs. In general, empirical results often depend on implicit assumptions, which should be articulated.
        \item The authors should reflect on the factors that influence the performance of the approach. For example, a facial recognition algorithm may perform poorly when image resolution is low or images are taken in low lighting. Or a speech-to-text system might not be used reliably to provide closed captions for online lectures because it fails to handle technical jargon.
        \item The authors should discuss the computational efficiency of the proposed algorithms and how they scale with dataset size.
        \item If applicable, the authors should discuss possible limitations of their approach to address problems of privacy and fairness.
        \item While the authors might fear that complete honesty about limitations might be used by reviewers as grounds for rejection, a worse outcome might be that reviewers discover limitations that aren't acknowledged in the paper. The authors should use their best judgment and recognize that individual actions in favor of transparency play an important role in developing norms that preserve the integrity of the community. Reviewers will be specifically instructed to not penalize honesty concerning limitations.
    \end{itemize}

\item {\bf Theory assumptions and proofs}
    \item[] Question: For each theoretical result, does the paper provide the full set of assumptions and a complete (and correct) proof?
    \item[] Answer: \answerNA{} 
    \item[] Justification: Our paper does not include theoretical results.
    \item[] Guidelines:
    \begin{itemize}
        \item The answer NA means that the paper does not include theoretical results. 
        \item All the theorems, formulas, and proofs in the paper should be numbered and cross-referenced.
        \item All assumptions should be clearly stated or referenced in the statement of any theorems.
        \item The proofs can either appear in the main paper or the supplemental material, but if they appear in the supplemental material, the authors are encouraged to provide a short proof sketch to provide intuition. 
        \item Inversely, any informal proof provided in the core of the paper should be complemented by formal proofs provided in appendix or supplemental material.
        \item Theorems and Lemmas that the proof relies upon should be properly referenced. 
    \end{itemize}

    \item {\bf Experimental result reproducibility}
    \item[] Question: Does the paper fully disclose all the information needed to reproduce the main experimental results of the paper to the extent that it affects the main claims and/or conclusions of the paper (regardless of whether the code and data are provided or not)?
    \item[] Answer: \answerYes{} 
    \item[] Justification: Our dataset and benchmarking code are made publicly available. We fully disclose our data generation process. Thus the results are readily reproducible.
    \item[] Guidelines:
    \begin{itemize}
        \item The answer NA means that the paper does not include experiments.
        \item If the paper includes experiments, a No answer to this question will not be perceived well by the reviewers: Making the paper reproducible is important, regardless of whether the code and data are provided or not.
        \item If the contribution is a dataset and/or model, the authors should describe the steps taken to make their results reproducible or verifiable. 
        \item Depending on the contribution, reproducibility can be accomplished in various ways. For example, if the contribution is a novel architecture, describing the architecture fully might suffice, or if the contribution is a specific model and empirical evaluation, it may be necessary to either make it possible for others to replicate the model with the same dataset, or provide access to the model. In general. releasing code and data is often one good way to accomplish this, but reproducibility can also be provided via detailed instructions for how to replicate the results, access to a hosted model (e.g., in the case of a large language model), releasing of a model checkpoint, or other means that are appropriate to the research performed.
        \item While NeurIPS does not require releasing code, the conference does require all submissions to provide some reasonable avenue for reproducibility, which may depend on the nature of the contribution. For example
        \begin{enumerate}
            \item If the contribution is primarily a new algorithm, the paper should make it clear how to reproduce that algorithm.
            \item If the contribution is primarily a new model architecture, the paper should describe the architecture clearly and fully.
            \item If the contribution is a new model (e.g., a large language model), then there should either be a way to access this model for reproducing the results or a way to reproduce the model (e.g., with an open-source dataset or instructions for how to construct the dataset).
            \item We recognize that reproducibility may be tricky in some cases, in which case authors are welcome to describe the particular way they provide for reproducibility. In the case of closed-source models, it may be that access to the model is limited in some way (e.g., to registered users), but it should be possible for other researchers to have some path to reproducing or verifying the results.
        \end{enumerate}
    \end{itemize}

\item {\bf Open access to data and code}
    \item[] Question: Does the paper provide open access to the data and code, with sufficient instructions to faithfully reproduce the main experimental results, as described in supplemental material?
    \item[] Answer: \answerYes{} 
    \item[] Justification: Our data and code are made publicly available.
    \item[] Guidelines:
    \begin{itemize}
        \item The answer NA means that paper does not include experiments requiring code.
        \item Please see the NeurIPS code and data submission guidelines (\url{https://nips.cc/public/guides/CodeSubmissionPolicy}) for more details.
        \item While we encourage the release of code and data, we understand that this might not be possible, so “No” is an acceptable answer. Papers cannot be rejected simply for not including code, unless this is central to the contribution (e.g., for a new open-source benchmark).
        \item The instructions should contain the exact command and environment needed to run to reproduce the results. See the NeurIPS code and data submission guidelines (\url{https://nips.cc/public/guides/CodeSubmissionPolicy}) for more details.
        \item The authors should provide instructions on data access and preparation, including how to access the raw data, preprocessed data, intermediate data, and generated data, etc.
        \item The authors should provide scripts to reproduce all experimental results for the new proposed method and baselines. If only a subset of experiments are reproducible, they should state which ones are omitted from the script and why.
        \item At submission time, to preserve anonymity, the authors should release anonymized versions (if applicable).
        \item Providing as much information as possible in supplemental material (appended to the paper) is recommended, but including URLs to data and code is permitted.
    \end{itemize}

\item {\bf Experimental setting/details}
    \item[] Question: Does the paper specify all the training and test details (e.g., data splits, hyperparameters, how they were chosen, type of optimizer, etc.) necessary to understand the results?
    \item[] Answer: \answerYes{} 
    \item[] Justification: We disclose models used, hyperparameters, and prompts in our paper and code.
    \item[] Guidelines:
    \begin{itemize}
        \item The answer NA means that the paper does not include experiments.
        \item The experimental setting should be presented in the core of the paper to a level of detail that is necessary to appreciate the results and make sense of them.
        \item The full details can be provided either with the code, in appendix, or as supplemental material.
    \end{itemize}

\item {\bf Experiment statistical significance}
    \item[] Question: Does the paper report error bars suitably and correctly defined or other appropriate information about the statistical significance of the experiments?
    \item[] Answer: \answerNo{} 
    \item[] Justification: We did not report any graphs that could include error bars. For our experimental results, this is a dataset paper, so we did not focus primarily on the results, but rather on the data.
    \item[] Guidelines:
    \begin{itemize}
        \item The answer NA means that the paper does not include experiments.
        \item The authors should answer "Yes" if the results are accompanied by error bars, confidence intervals, or statistical significance tests, at least for the experiments that support the main claims of the paper.
        \item The factors of variability that the error bars are capturing should be clearly stated (for example, train/test split, initialization, random drawing of some parameter, or overall run with given experimental conditions).
        \item The method for calculating the error bars should be explained (closed form formula, call to a library function, bootstrap, etc.)
        \item The assumptions made should be given (e.g., Normally distributed errors).
        \item It should be clear whether the error bar is the standard deviation or the standard error of the mean.
        \item It is OK to report 1-sigma error bars, but one should state it. The authors should preferably report a 2-sigma error bar than state that they have a 96\% CI, if the hypothesis of Normality of errors is not verified.
        \item For asymmetric distributions, the authors should be careful not to show in tables or figures symmetric error bars that would yield results that are out of range (e.g. negative error rates).
        \item If error bars are reported in tables or plots, The authors should explain in the text how they were calculated and reference the corresponding figures or tables in the text.
    \end{itemize}

\item {\bf Experiments compute resources}
    \item[] Question: For each experiment, does the paper provide sufficient information on the computer resources (type of compute workers, memory, time of execution) needed to reproduce the experiments?
    \item[] Answer: \answerYes{} 
    \item[] Justification: All experiments are run on AWS and model inference is conducted using the AWS Bedrock API. Thus the details of compute needs are not readily available.
    \item[] Guidelines:
    \begin{itemize}
        \item The answer NA means that the paper does not include experiments.
        \item The paper should indicate the type of compute workers CPU or GPU, internal cluster, or cloud provider, including relevant memory and storage.
        \item The paper should provide the amount of compute required for each of the individual experimental runs as well as estimate the total compute. 
        \item The paper should disclose whether the full research project required more compute than the experiments reported in the paper (e.g., preliminary or failed experiments that didn't make it into the paper). 
    \end{itemize}
    
\item {\bf Code of ethics}
    \item[] Question: Does the research conducted in the paper conform, in every respect, with the NeurIPS Code of Ethics \url{https://neurips.cc/public/EthicsGuidelines}?
    \item[] Answer: \answerYes{} 
    \item[] Justification: Our paper does not violate the ethics code in any way.
    \item[] Guidelines:
    \begin{itemize}
        \item The answer NA means that the authors have not reviewed the NeurIPS Code of Ethics.
        \item If the authors answer No, they should explain the special circumstances that require a deviation from the Code of Ethics.
        \item The authors should make sure to preserve anonymity (e.g., if there is a special consideration due to laws or regulations in their jurisdiction).
    \end{itemize}

\item {\bf Broader impacts}
    \item[] Question: Does the paper discuss both potential positive societal impacts and negative societal impacts of the work performed?
    \item[] Answer: \answerYes{} 
    \item[] Justification: The paper discusses the positive impact of making IaC more accessible through natural language interfaces. We do not see any major negative impacts. 
    \item[] Guidelines:
    \begin{itemize}
        \item The answer NA means that there is no societal impact of the work performed.
        \item If the authors answer NA or No, they should explain why their work has no societal impact or why the paper does not address societal impact.
        \item Examples of negative societal impacts include potential malicious or unintended uses (e.g., disinformation, generating fake profiles, surveillance), fairness considerations (e.g., deployment of technologies that could make decisions that unfairly impact specific groups), privacy considerations, and security considerations.
        \item The conference expects that many papers will be foundational research and not tied to particular applications, let alone deployments. However, if there is a direct path to any negative applications, the authors should point it out. For example, it is legitimate to point out that an improvement in the quality of generative models could be used to generate deepfakes for disinformation. On the other hand, it is not needed to point out that a generic algorithm for optimizing neural networks could enable people to train models that generate Deepfakes faster.
        \item The authors should consider possible harms that could arise when the technology is being used as intended and functioning correctly, harms that could arise when the technology is being used as intended but gives incorrect results, and harms following from (intentional or unintentional) misuse of the technology.
        \item If there are negative societal impacts, the authors could also discuss possible mitigation strategies (e.g., gated release of models, providing defenses in addition to attacks, mechanisms for monitoring misuse, mechanisms to monitor how a system learns from feedback over time, improving the efficiency and accessibility of ML).
    \end{itemize}
    
\item {\bf Safeguards}
    \item[] Question: Does the paper describe safeguards that have been put in place for responsible release of data or models that have a high risk for misuse (e.g., pretrained language models, image generators, or scraped datasets)?
    \item[] Answer: \answerNA{} 
    \item[] Justification: There are not such risks.
    \item[] Guidelines:
    \begin{itemize}
        \item The answer NA means that the paper poses no such risks.
        \item Released models that have a high risk for misuse or dual-use should be released with necessary safeguards to allow for controlled use of the model, for example by requiring that users adhere to usage guidelines or restrictions to access the model or implementing safety filters. 
        \item Datasets that have been scraped from the Internet could pose safety risks. The authors should describe how they avoided releasing unsafe images.
        \item We recognize that providing effective safeguards is challenging, and many papers do not require this, but we encourage authors to take this into account and make a best faith effort.
    \end{itemize}

\item {\bf Licenses for existing assets}
    \item[] Question: Are the creators or original owners of assets (e.g., code, data, models), used in the paper, properly credited and are the license and terms of use explicitly mentioned and properly respected?
    \item[] Answer: \answerYes{} 
    \item[] Justification: All datasets used as source data are properly credited.
    \item[] Guidelines:
    \begin{itemize}
        \item The answer NA means that the paper does not use existing assets.
        \item The authors should cite the original paper that produced the code package or dataset.
        \item The authors should state which version of the asset is used and, if possible, include a URL.
        \item The name of the license (e.g., CC-BY 4.0) should be included for each asset.
        \item For scraped data from a particular source (e.g., website), the copyright and terms of service of that source should be provided.
        \item If assets are released, the license, copyright information, and terms of use in the package should be provided. For popular datasets, \url{paperswithcode.com/datasets} has curated licenses for some datasets. Their licensing guide can help determine the license of a dataset.
        \item For existing datasets that are re-packaged, both the original license and the license of the derived asset (if it has changed) should be provided.
        \item If this information is not available online, the authors are encouraged to reach out to the asset's creators.
    \end{itemize}

\item {\bf New assets}
    \item[] Question: Are new assets introduced in the paper well documented and is the documentation provided alongside the assets?
    \item[] Answer: \answerYes{} 
    \item[] Justification: Our paper provides new data assets (it is a dataset paper) and we properly document the data provided.
    \item[] Guidelines:
    \begin{itemize}
        \item The answer NA means that the paper does not release new assets.
        \item Researchers should communicate the details of the dataset/code/model as part of their submissions via structured templates. This includes details about training, license, limitations, etc. 
        \item The paper should discuss whether and how consent was obtained from people whose asset is used.
        \item At submission time, remember to anonymize your assets (if applicable). You can either create an anonymized URL or include an anonymized zip file.
    \end{itemize}

\item {\bf Crowdsourcing and research with human subjects}
    \item[] Question: For crowdsourcing experiments and research with human subjects, does the paper include the full text of instructions given to participants and screenshots, if applicable, as well as details about compensation (if any)? 
    \item[] Answer: \answerNo{} 
    \item[] Justification: The paper involved a limited internal human review, so there was no additional compensation provided. Review was based on expert knowledge and did not include detailed instructions to annotators.
    \item[] Guidelines:
    \begin{itemize}
        \item The answer NA means that the paper does not involve crowdsourcing nor research with human subjects.
        \item Including this information in the supplemental material is fine, but if the main contribution of the paper involves human subjects, then as much detail as possible should be included in the main paper. 
        \item According to the NeurIPS Code of Ethics, workers involved in data collection, curation, or other labor should be paid at least the minimum wage in the country of the data collector. 
    \end{itemize}

\item {\bf Institutional review board (IRB) approvals or equivalent for research with human subjects}
    \item[] Question: Does the paper describe potential risks incurred by study participants, whether such risks were disclosed to the subjects, and whether Institutional Review Board (IRB) approvals (or an equivalent approval/review based on the requirements of your country or institution) were obtained?
    \item[] Answer: \answerNo{} 
    \item[] Justification: A very limited, intermal human evaluation of our generated data was conducted. This did not necessitate IRB approval.
    \item[] Guidelines:
    \begin{itemize}
        \item The answer NA means that the paper does not involve crowdsourcing nor research with human subjects.
        \item Depending on the country in which research is conducted, IRB approval (or equivalent) may be required for any human subjects research. If you obtained IRB approval, you should clearly state this in the paper. 
        \item We recognize that the procedures for this may vary significantly between institutions and locations, and we expect authors to adhere to the NeurIPS Code of Ethics and the guidelines for their institution. 
        \item For initial submissions, do not include any information that would break anonymity (if applicable), such as the institution conducting the review.
    \end{itemize}

\item {\bf Declaration of LLM usage}
    \item[] Question: Does the paper describe the usage of LLMs if it is an important, original, or non-standard component of the core methods in this research? Note that if the LLM is used only for writing, editing, or formatting purposes and does not impact the core methodology, scientific rigorousness, or originality of the research, declaration is not required.
    \item[] Answer: \answerYes{} 
    \item[] Justification: We fully disclose our use of LLMs in our data generation pipeline. For the paper itself, LLMs were used for editing but not for content creation.
    \item[] Guidelines:
    \begin{itemize}
        \item The answer NA means that the core method development in this research does not involve LLMs as any important, original, or non-standard components.
        \item Please refer to our LLM policy (\url{https://neurips.cc/Conferences/2025/LLM}) for what should or should not be described.
    \end{itemize}

\end{enumerate}

\end{document}